\documentclass[10pt]{article}
\usepackage[twoside,pdftex,papersize={210mm,280mm}, total={16.8cm,23.6cm}, left=1.5cm, top=1.5cm, right=1.5cm, bottom=2.5cm, includehead=true]{geometry}
\usepackage{amssymb}
\usepackage[affil-it]{authblk}
\usepackage[fleqn]{amsmath}
\usepackage{lineno}
\usepackage{epsfig}
\usepackage{subfigure}
\usepackage{hyperref}
 
\begin{document}
 
\title{Identification of circles from datapoints using Gaussian sums}

\author{T.~Alexopoulos$^{1}$, G.~Iakovidis$^{1,2}$, S.~Leontsinis$^{1,2}$\footnote{stefanos.leontsinis@cern.ch}\ , K.~Ntekas$^{1,2}$ and V.~Polychronakos$^{2}$}

\affil{$^1$National Technical University of Athens}
\affil{$^2$Brookhaven National Laboratory}

\maketitle
 
\begin{abstract}
We present a pattern recognition method which use datapoints on a plane and estimates the parameters of a circle. MC data are generated in order to test the method's efficiency over noise hits, uncertainty in the hits positions and number of datapoints. The scenario were the hits from a quadrant of the circle are missing is also considered. The method proposed is proven to be robust, accurate and very efficient.
\end{abstract}
 
\section{Introduction}
Various types of detectors require algorithms that handle number of datapoints in a plane and reconstruct the parameters of a circle, \cite{cherenkov_1},\cite{cherenkov_2},\cite{cherenkov_3}. These algorithms need to be robust against noise hits and analyse big amount of data relatively fast. Finally, the performance of the algorithm needs as independent as possible with the resolution of the detector.

In a recent paper \cite{legendre_paper} we examined the use of the Legendre transform for the determination of a circle's characteristics. In this paper we revisit this problem and describe in detail an alternative method, based on Gaussian sums, proposed for the reconstruction of the circle's center $(x_0,y_0)$ and radius $R$ from a given set of datapoints.
 
\section{Description of the Methods}
\label{sec:model_description}

\begin{figure}[h]\centering
         \includegraphics[scale=.18]{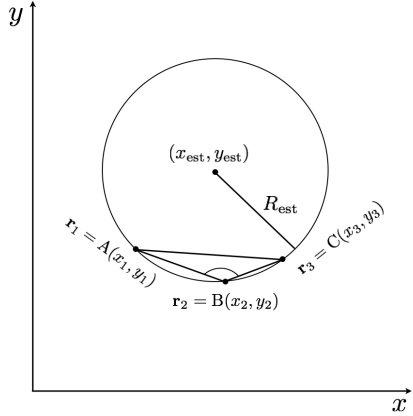}
        \caption{Representation of a circle with center $(x_\mathrm{est},y_\mathrm{est})$ and radius $R_\mathrm{est}$.}
	\label{fig:circle_circle}
\end{figure}

Following the notation and train of though of \cite{legendre_paper}, using $3$ datapoints (${\bf r_1}=A(x_1,y_1)$, ${\bf r_2}=B(x_2,y_2)$ and ${\bf r_3}=C(x_3,y_3)$), we can have a first estimation of the center $(x_\mathrm{est}, y_{\mathrm{est}})$ and the radius, $R_\mathrm{est}$, of the circle (see figure \ref{fig:circle_circle}). 

These estimates are given by:

\begin{equation*}\begin{aligned}
x_\mathrm{est} &= \frac{m_t m_r(y_3-y_1) + m_r(x_2+x_3) - m_t(x_1+x_2)}{2(m_r-m_t)}
\\
y_\mathrm{est} &= -\frac{1}{m_r}\left( x_\mathrm{est} - \frac{x_1+x_2}{2} \right) + \frac{y_1+y_2}{2}
\\
R_\mathrm{est} &= \sqrt{(x_\mathrm{est}-x_1)^2 + (y_\mathrm{est} - y_1)^2}
\end{aligned}\end{equation*}
where $m_t=(y_3-y_2)/(x_3-x_2)$ and $m_r=(y_2-y_1)/(x_2-x_1)$ the slopes of the lines connecting $AB$ and $BC$.
  
\subsection{Gaussian Sum}
\label{sec:model_description_first}
Having $n$ given datapoints/measurements, $k=\left({n \atop 3}\right)$ different centers and radii can be estimated. Using the formulas described above the global sums are constructed
 
\begin{equation*}\begin{aligned}
& G(x)=\sum_{i=1}^{k}\frac{1}{\sqrt{2\pi}\sigma_x}e^{- (x-x_\mathrm{est}^i)^2 / 2\sigma_x^2}
\\
& G(y)=\sum_{i=1}^{k}\frac{1}{\sqrt{2\pi}\sigma_y}e^{- (y-y_\mathrm{est}^i)^2 / 2\sigma_y^2}
\\
& G(R)=\sum_{i=1}^{k}\frac{1}{\sqrt{2\pi}\sigma_R}e^{- (R-R_\mathrm{est}^i)^2 / 2\sigma_R^2}
\end{aligned}\end{equation*}
where $x_\mathrm{est}^i$, $y_\mathrm{est}^i$ and $R_\mathrm{est}^i$ are the circle's parameters for the $i^\mathrm{th}$ triplet of datapoints that defines the circle. The standard deviations of the gaussians are set of a fraction of the geometrical scale of the problem. In the studies below $\sigma_x=\sigma_y=\sigma_R=0.1$ have been used. The reason of using gaussian functions in the definition of the global sums (see above) instead of using the basic $\chi^2$ test functions, is based on the fact that noise type datapoints will give values of the parameters far away from the right answer. These estimated parameters ($x_\mathrm{est}^i$, $y_\mathrm{est}^i$ and $R_\mathrm{est}^i$) will contribute very little to the global sums. On the contrary, the right values of the estimated parameters will give the maximum contribution.
 
\begin{figure}[h]\centering
        \centering
        \subfigure[] {
                \includegraphics[scale=.18]{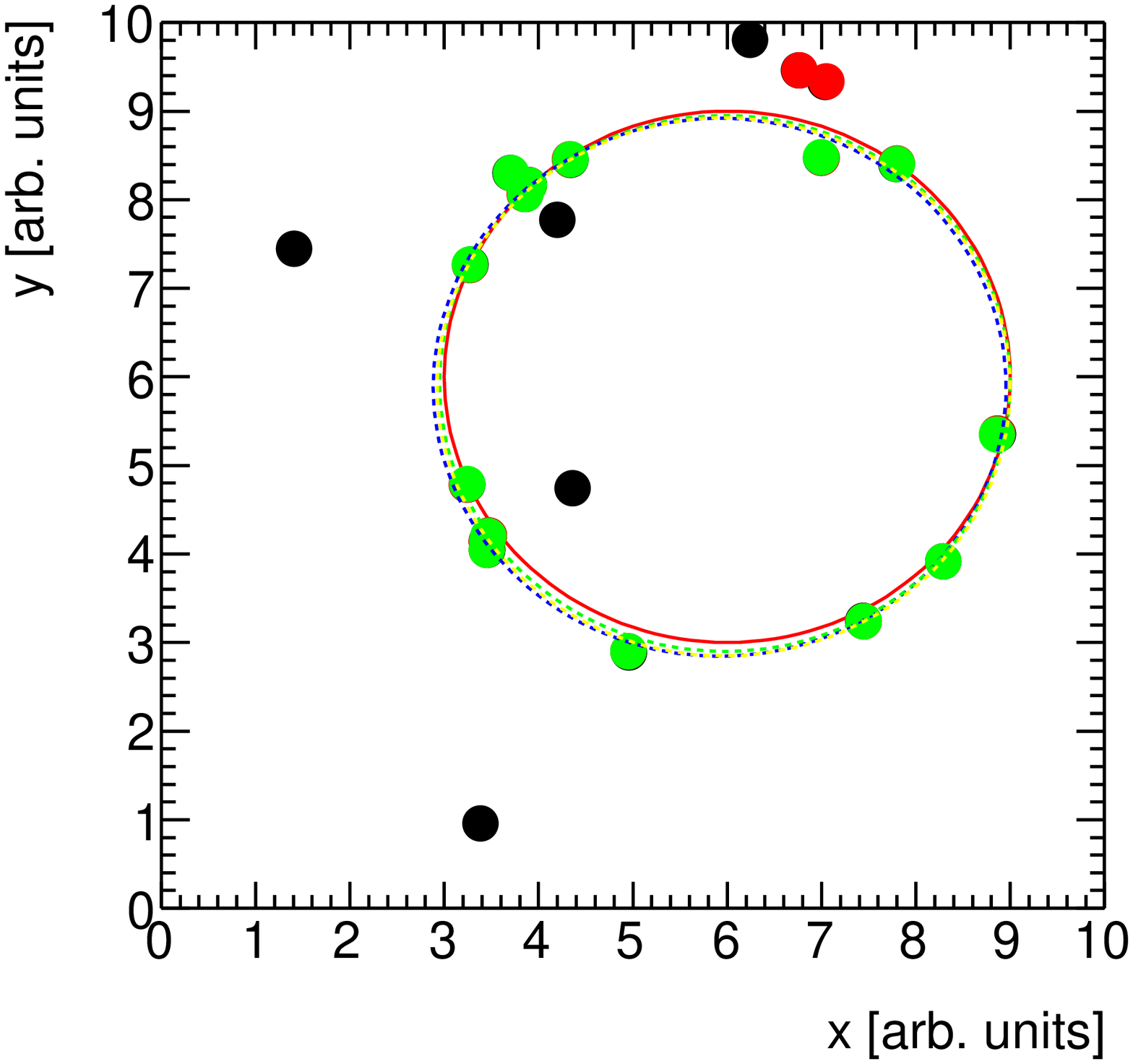}
                \label{fig:one_circle}
        }
        \subfigure[] {
                \includegraphics[scale=.18]{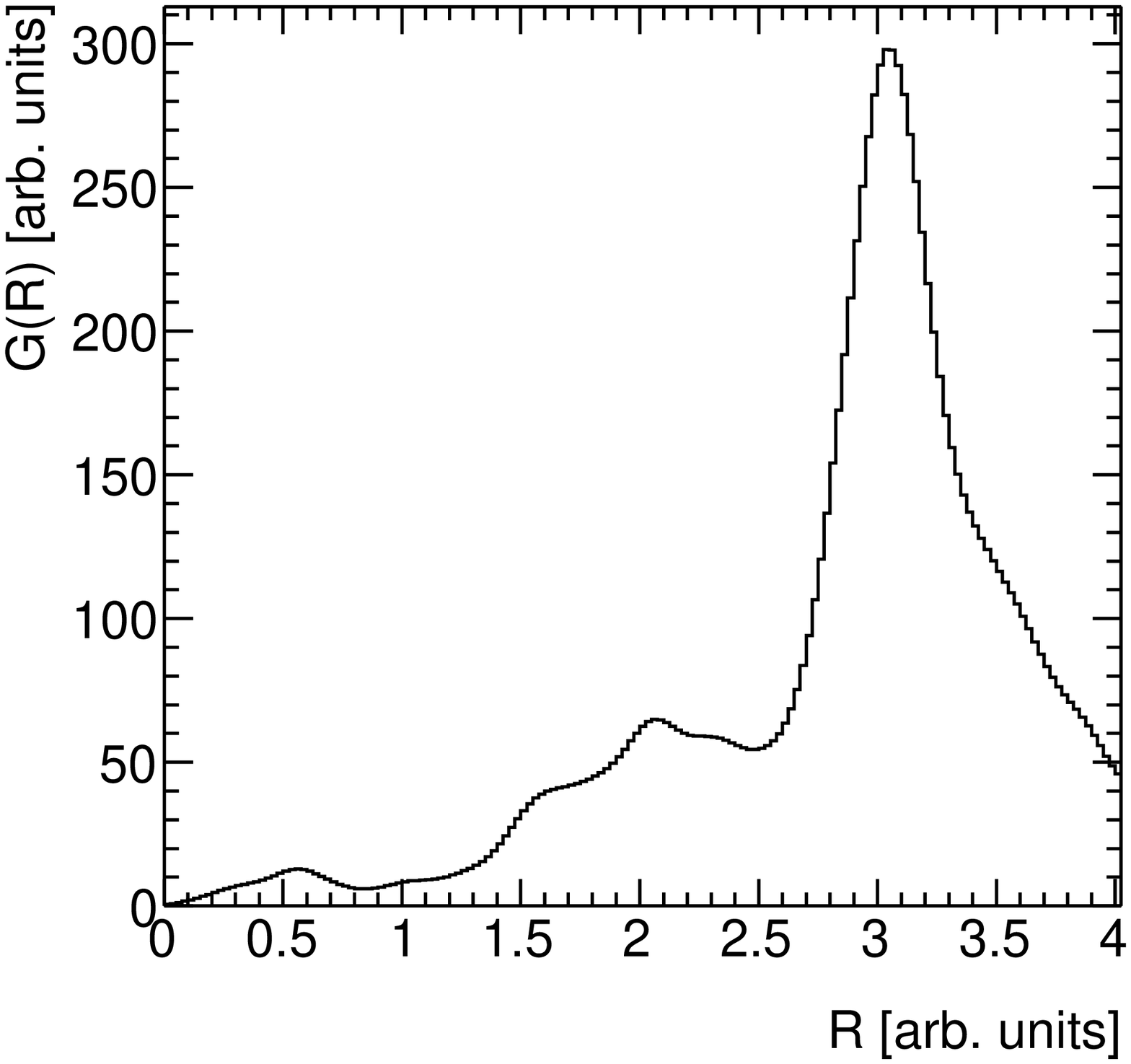}
                \label{fig:one_circle_R_gauss}
        }
        \subfigure[] {
                \includegraphics[scale=.18]{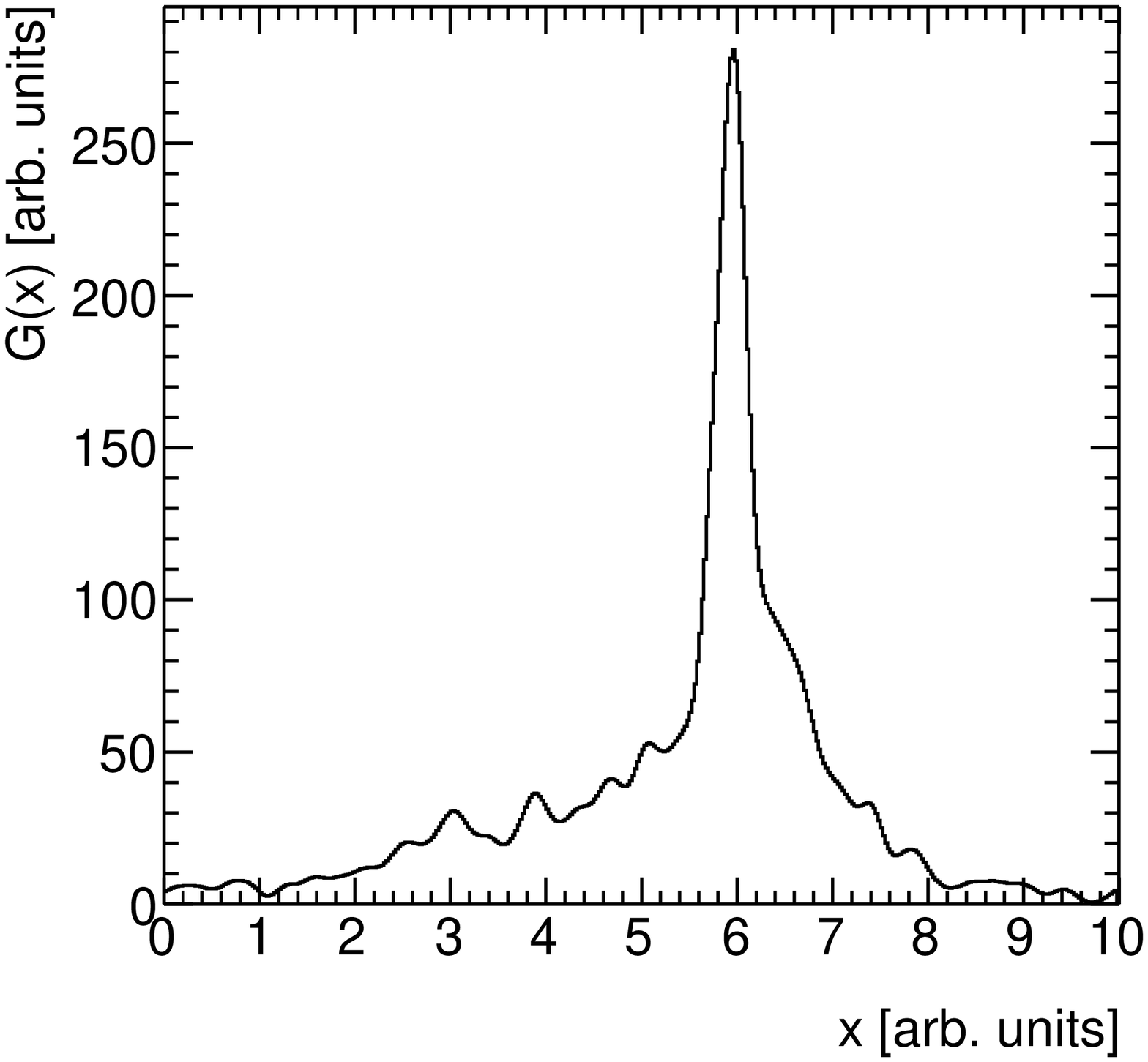}
                \label{fig:one_circle_x_gauss}
        }
        \subfigure[] {
                \includegraphics[scale=.18]{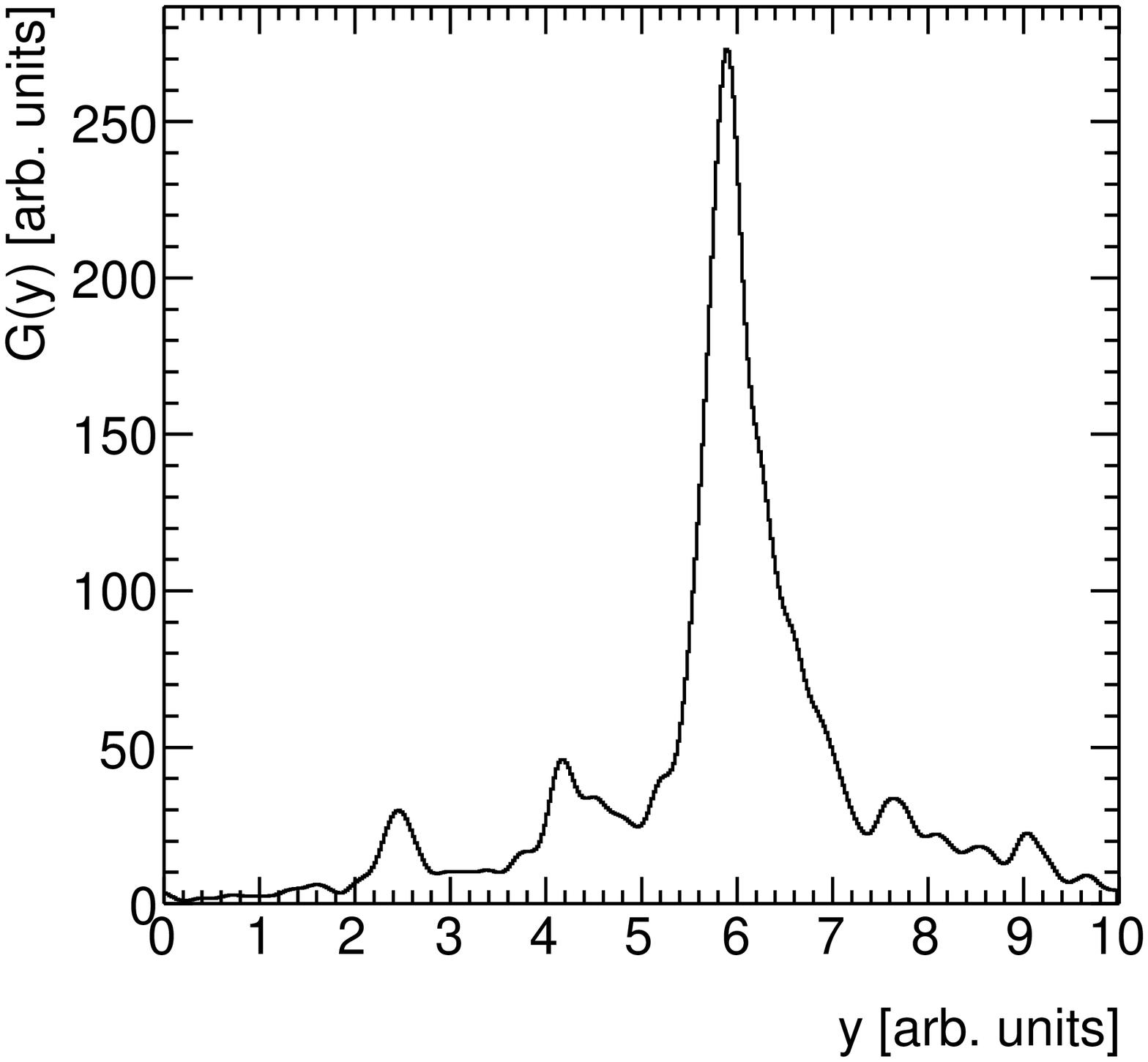}
                \label{fig:one_circle_y_gauss}
        }
        \caption{(a) The red datapoints originate from the circle (red line) having received a smearing of 10\%. The datapoints that missed the radius $R$ (noise hits - black datapoints) are on a 50\% percentage of the circle's datapoints. Red circle is the original circle used to generate the datapoints, blue is reconstructed by using $\chi^2$ minimization, green is reconstructed by using the Legendre technique and yellow using the gaussian sum. (b) Example of the $G(R)$ function for the reconstruction of the radius $R$. (c) Example of the $G(x)$ function for the reconstruction of the $x_0$. (d) Example of the $G(y)$ function for the reconstruction of the $y_0$.}
        \label{fig:gaussians_sum_example}
\end{figure}

For each event the functions described above (see Figure \ref{fig:gaussians_sum_example}) are constructed and the value of the bin with the maximum number of entries is chosen. The value of these bins are the estimate of the $(x_\mathrm{est},y_\mathrm{est})$ and $R_\mathrm{est}$ of the circle.
 
\subsection{Extraction of the Circle Parameters}
\label{sec:sub_extracion}

The extraction of the circle parameters using the first method (see section \ref{sec:model_description_first}) is straightforward: the bin with the maximum number of entries per event in each $G(x)$, $G(y)$ and $G(R)$ gives the estimate of the center's coordinates and the radius of the circle. An example event can be seen in Figure \ref{fig:gaussians_sum_example}. In Figure \ref{fig:one_circle} the circle used to generate the datapoints is shown. The $G(R)$, $G(x)$ and $G(y)$ distributions for the circle parameters can be seen in Figures \ref{fig:one_circle_R_gauss}, \ref{fig:one_circle_x_gauss} and \ref{fig:one_circle_y_gauss} respectively. $G(x)$ and $G(y)$ for $x_\mathrm{est}$ and $y_\mathrm{est}$ peak at $6$ where the center of the circle is and $G(R)$ peaks at $R_\mathrm{est}=3$ indicating the actual radius of the circle.

\section{Performance Studies}
\label{sec:application}
The algorithm is tested using simulated experiments, creating datapoints ($n=8-20$) that belong to a circle. The robustness of the algorithms was tested by inserting gaussian uncertainties in both the $x$ and $y$ coordinates of the datapoints/measurements and random noise hits/outliers.
 
A special scenario was considered, where one quadrant of the detector was deactivated (results are summarized in table \ref{tab:Results_no_quad}).

\subsection{One Circle Case}
\label{sec:one_circle_case}
The center of the circle was chosen to be at $(4,5)$ with a radius of $2$ (arbitrary units). Using this circle, $n$ datapoints ($n= 8,\, 10,\, 15$ and $20$) are randomly created. We smear both the $x_i$ and $y_i$ positions of the datapoints, independently on $x$ and $y$, by $5\%$, $10\%$ or $15\%$ of the circle's radius $R$. Apart from the uncertainty introduced in the position of the datapoints, noise hits in certain percentage of the datapoints that originate from the circle are included. The noise hits are randomly generated in the ranges $0\leq x\leq 10$ and $0\leq y\leq 10$. An event generated by this toy MC can be seen in Figure \ref{fig:one_circle}. The red and green datapoints ($n=15$) are generated from the circle and the black datapoints are the noise hits ($50\%$ noise level is used in this event). The positions of all the generated datapoints $(x_i,y_i)$ are smeared by $10\%$ of its radius $R$.

The results of the methods described in section \ref{sec:model_description} are summarized in Table \ref{tab:Results}. The algorithm proposed for the extraction of the circle's parameters work well under all test scenarios.
 
\begin{table}[h!]
\begin{center}
\caption{\normalsize Results of the algorithms described in section \ref{sec:sub_extracion}. We use $n=10$ datapoints, generated in three quadrants of the circle, three different percentages of position uncertainty ($5\%$, $10\%$ and $15\%$ of the radius $R$) and five additional noise hits. The original values of the circle's parameters can be found in the first row.}
\vspace{1ex}
\begin{tabular}{c c c c}
\hline\hline
Original Values & $R=3$ & $x=6$ & $y=6$ \\\hline
Uncertainty [\%]& 5 & 10 & 15 \\ 
R              & $3.01 \pm 0.00$ & $3.03 \pm 0.00$ & $3.04 \pm 0.01$ \\ 
x              & $6.00 \pm 0.00$ & $5.99 \pm 0.01$ & $6.00 \pm 0.02$ \\ 
y              & $6.00 \pm 0.00$ & $5.99 \pm 0.01$ & $5.98 \pm 0.01$ \\ \hline\hline
\end{tabular}
\label{tab:Results_no_quad}
\end{center}
\end{table}

 \begin{table}[h!]\scriptsize
\begin{center}
\caption{\normalsize Results of the algorithms described in section \ref{sec:sub_extracion}. We try four different sets of number of datapoints ($n=8,10,15,20$), three different percentages of position uncertainty ($5\%$, $10\%$ and $15\%$ of the radius $R$) and three levels of noise hits ($0\%,\ 25\%$ and $50\%$ of the actual datapoints). The original values of the circle's parameters can be found in the first row.}
\vspace{1ex}
\begin{tabular}{|c|c|c|c|c|c|c|c|c|c|c|c|}
\hline
Original Values & \multicolumn{3}{|c|}{$R=2$} & \multicolumn{3}{|c|}{$x=5$}  &\multicolumn{3}{|c|}{$y=4$}  \\\hline
\multicolumn{10}{c}{$n=8$}\\
\hline\hline
Noise [\%] & \multicolumn{3}{|c|}{0}                                                    & \multicolumn{3}{|c|}{25}                                                                                                      & \multicolumn{3}{|c|}{50}  \\\hline
Uncertainty [\%]&      5                                        &                               10      &               15                    &          5                                  &           10                                &             15                                      &               5                                               &               10                              &               15 \\ \hline
R             & $2.00 \pm 0.00$ & $2.01 \pm 0.00$ & $2.02 \pm 0.01$ & $2.01 \pm 0.00$ & $2.02 \pm 0.00$ & $2.02 \pm 0.01$ & $2.01 \pm 0.00$ & $2.02 \pm 0.00$ & $2.04 \pm 0.01$ \\ \hline
x              & $5.00 \pm 0.00$ & $5.00 \pm 0.01$ & $5.00 \pm 0.01$ & $5.00 \pm 0.00$ & $5.00 \pm 0.01$ & $5.00 \pm 0.01$ & $5.00 \pm 0.00$ & $5.00 \pm 0.01$ & $5.02 \pm 0.01$ \\ \hline
y              & $4.00 \pm 0.00$ & $4.00 \pm 0.01$ & $4.02 \pm 0.01$ & $4.00 \pm 0.00$ & $4.00 \pm 0.01$ & $4.01 \pm 0.01$ & $4.00 \pm 0.00$ & $4.01 \pm 0.01$ & $4.01 \pm 0.01$ \\ \hline
 
\multicolumn{10}{c}{$n=10$}\\
\hline\hline
Noise [\%] & \multicolumn{3}{|c|}{0}                                                                      & \multicolumn{3}{|c|}{25}                                                                                                    & \multicolumn{3}{|c|}{50}  \\\hline
Uncertainty [\%]&      5                                        &               10          &           15                    &          5                                  &           10                                &             15                                      &               5                                       &               10                                      &               15 \\ \hline
R             & $2.01 \pm 0.00$ & $2.01 \pm 0.00$ & $2.01 \pm 0.01$ & $2.00 \pm 0.00$ & $2.02 \pm 0.00$ & $2.03 \pm 0.01$ & $2.01 \pm 0.00$ & $2.03 \pm 0.00$ & $2.04 \pm 0.01$ \\ \hline
x              & $5.00 \pm 0.00$ & $5.00 \pm 0.00$ & $5.00 \pm 0.00$ & $5.00 \pm 0.00$ & $5.01 \pm 0.01$ & $4.99 \pm 0.01$ & $5.00 \pm 0.00$ & $5.00 \pm 0.01$ & $4.99 \pm 0.01$ \\ \hline
y              & $4.00 \pm 0.00$ & $4.01 \pm 0.00$ & $4.00 \pm 0.01$ & $4.00 \pm 0.00$ & $4.00 \pm 0.00$ & $4.00 \pm 0.01$ & $4.00 \pm 0.00$ & $4.00 \pm 0.01$ & $4.00 \pm 0.01$ \\ \hline
\multicolumn{10}{c}{$n=15$}\\
\hline\hline
Noise [\%] & \multicolumn{3}{|c|}{0}                                                                  & \multicolumn{3}{|c|}{25}                                                                                                        & \multicolumn{3}{|c|}{50}  \\\hline
Uncertainty [\%]&      5                                        &               10          &           15                    &          5                                  &           10                                &             15                                      &               5                                               &               10                              &               15 \\ \hline
R             & $2.01 \pm 0.00$ & $2.01 \pm 0.00$ & $2.01 \pm 0.00$ & $2.01 \pm 0.02$ & $2.01 \pm 0.00$ & $2.02 \pm 0.00$ & $2.01 \pm 0.00$ & $2.02 \pm 0.00$ & $2.02 \pm 0.00$ \\ \hline
x              & $5.00 \pm 0.00$ & $5.00 \pm 0.00$ & $5.00 \pm 0.01$ & $5.00 \pm 0.00$ & $5.00 \pm 0.00$ & $5.00 \pm 0.01$ & $5.00 \pm 0.00$ & $5.01 \pm 0.00$ & $4.99 \pm 0.01$ \\ \hline
y              & $4.00 \pm 0.00$ & $4.00 \pm 0.00$ & $4.01 \pm 0.01$ & $4.00 \pm 0.00$ & $4.00 \pm 0.00$ & $4.00 \pm 0.01$ & $4.00 \pm 0.00$ & $4.00 \pm 0.00$ & $4.00 \pm 0.01$ \\ \hline
 
\multicolumn{10}{c}{$n=20$}\\
\hline\hline
Noise [\%] & \multicolumn{3}{|c|}{0}                                                            & \multicolumn{3}{|c|}{25}                                                                                                      & \multicolumn{3}{|c|}{50}  \\\hline
Uncertainty [\%]&      5                                        &               10          &           15                    &          5                                  &           10                                &             15                                      &               5                                       &                       10                              &               15 \\ \hline
R             & $2.01 \pm 0.01$ & $2.01 \pm 0.00$ & $2.02 \pm 0.00$ & $2.01 \pm 0.01$ & $2.01 \pm 0.00$ & $2.03 \pm 0.00$ & $2.02 \pm 0.02$ & $2.02 \pm 0.00$ & $2.03 \pm 0.00$ \\ \hline
x             & $5.00 \pm 0.00$ & $5.00 \pm 0.00$ & $4.99 \pm 0.00$ & $5.00 \pm 0.00$ & $5.00 \pm 0.00$ & $5.00 \pm 0.00$ & $5.00 \pm 0.00$ & $5.00 \pm 0.00$ & $5.00 \pm 0.01$ \\ \hline
y             & $4.00 \pm 0.01$ & $4.00 \pm 0.00$ & $4.00 \pm 0.01$ & $4.01 \pm 0.00$ & $4.00 \pm 0.00$ & $4.00 \pm 0.01$ & $4.00 \pm 0.00$ & $4.00 \pm 0.00$ & $4.00 \pm 0.01$ \\ \hline
\end{tabular}
\label{tab:Results}
\end{center}
\end{table}

\normalsize
 
\section{Conclusions}
We examined the use of Gaussian sums for the reconstruction of a circle's center and radius, using its datapoints. The algorithm proposed proved to work well under all test scenarios (noise hits, inactive part of the detector, hit position uncertainties) and provide fast and accurate results.

\section*{Acknowledgments}
The research leading to these results has received funding from the FP7-INFRASTRUCTURES-2010, AIDA: Advanced European Infrastructures for Detectors at  Accelerators, Grant Agreement No: 262025 and the European Union (European Social Fund ESF) and the Greek national funds through the Operational Program "Education and Lifelong Learning" of the National Strategic Reference Framework (NSRF) 2007-1013. ARISTEIA-1893-ATLAS MICROMEGAS. The support is gratefully acknowledged.

\end{document}